\documentclass[12pt]{article}
\usepackage{graphicx}

\usepackage{geometry}

\usepackage{orcidlink}
\usepackage{cite}
\usepackage[initials,citation-order,nobysame]{amsrefs}

\BibSpec{article}{%choose what to show in the bibliography
+{} {\PrintAuthors} {author}
+{.} { } {part}
+{:} { \textit} {subtitle}
+{,} { \PrintContributions} {contribution}
+{.} { \PrintPartials} {partial}
+{,} { } {journal}
+{} { \textbf} {volume}
+{} { \PrintDatePV} {date}
+{,} { \issuetext} {number}
+{,} { \eprintpages} {pages}
+{,} { } {status}
+{,} { \PrintDOI} {doi}
+{,} { \eprint} {eprint}
+{} { \parenthesize} {language}
+{} { \PrintTranslation} {translation}
+{;} { \PrintReprint} {reprint}
+{.} { } {note}
+{.} {} {transition}
+{} {\SentenceSpace \PrintReviews} {review}
}

\usepackage{verbatim}

\usepackage{float}
\usepackage{subfloat}

\usepackage{mathtools}

\usepackage[font=scriptsize]{caption}
\usepackage{subcaption}

\usepackage{ragged2e}
\usepackage{amsmath}
\usepackage{amsfonts}
\usepackage{amssymb}
\usepackage{graphicx}
\usepackage{slashed}
\usepackage{bm}
\usepackage{color}
\usepackage{epsf}
 \usepackage{orcidlink}
\usepackage{psfrag}
\usepackage{appendix}

\usepackage[parfill]{parskip}
\usepackage{hyperref}
\hypersetup{
  colorlinks   = true, %Colours links instead of ugly boxes
  urlcolor     = black, %Colour for external hyperlinks
  linkcolor    = black, %Colour of internal links
  citecolor   = red %Colour of citations
}

\newcommand{\bq}{\bar{q}} % \bq = (q + q')/2
\newcommand{\bp}{\bar{p}} % \bp = (p + p')/2
\newcommand{\thetaL}{\theta_\ell} % polar angle of muon's 3-momentum in leptons' CM frame
\newcommand{\phiL}{\phi_\ell} % azimuthal angle of muon's 3-momentum in leptons' CM frame
\newcommand{\M}{\mathcal{M}} % amplitude
 
\newcommand{\cffh}{\mathcal{H}} % CFF H
\newcommand{\cffe}{\mathcal{E}} % CFF E
\newcommand{\cffht}{\mathcal{\widetilde{H}}} % CFF Htilde
\newcommand{\cffet}{\mathcal{\widetilde{E}}} % CFF Etilde

\newcommand{\Jt}{\mathcal{J}^{(2)}}
\newcommand{\Jofplus}{\mathcal{J}^{(1, 5)+}}
\newcommand{\Jtfplus}{\mathcal{J}^{(2, 5)+}}

\newcommand{\phiLBDP}{\phi_{\ell, \mathrm{BDP}}}
\newcommand{\thetaLBDP}{\theta_{\ell, \mathrm{BDP}}}
\newcommand{\OmegaLBDP}{\Omega_{\ell, \mathrm{BDP}}}

\newcommand\pubdate{\today}

\def\Title#1{\begin{center} {\Large #1 } \end{center}}
\def\Author#1{\begin{center}{ \sc #1} \end{center}}
\def\Address#1{\begin{center}{ \it #1} \end{center}}

\newcommand\pubblock{\rightline{\begin{tabular}{l}
         \pubdate  \end{tabular}}}
\newenvironment{Abstract}{\begin{quotation}  }{\end{quotation}}
\newenvironment{Presented}{\begin{quotation} \begin{center} 
             PRESENTED BY V.M.F.~AT\end{center}\bigskip 
      \begin{center}\begin{large}}{\end{large}\end{center} \end{quotation}}
\begin{document}
\begin{titlepage}
 \pubblock
\vfill
\Title{{\bf Double DVCS as a window to the complete mapping of GPDs}}
\vfill
\Author{K.~Deja$^\dagger$\,\orcidlink{0000-0002-9083-2382}, V.~Mart\'inez-Fern\'andez$^\dagger$\,\orcidlink{0000-0002-0581-7154}, B.~Pire$^*$\,\orcidlink{0000-0003-4882-7800}, P.~Sznajder$^\dagger$\,\orcidlink{0000-0002-2684-803X}, J.~Wagner$^\dagger$\,\orcidlink{0000-0001-8335-7096}}
\Address{${}^\dagger$National Centre for Nuclear Research (NCBJ), 02-093 Warsaw, Poland\\\vspace{2mm}${}^*$Centre de Physique Th\'eorique, CNRS, \'Ecole Polytechnique, I.P. Paris, 91128 Palaiseau, France}
\vfill
\begin{Abstract}
Double deeply virtual Compton scattering (DDVCS) is the process where an electron scatters off a nucleon and produces a lepton pair. The main advantage of this process in contrast with deeply virtual and timelike Compton scatterings (DVCS and TCS) is the possibility of directly measuring GPDs for $x\neq\pm\xi$ at leading order in $\alpha_s$ (LO). We present a new calculation of the DDVCS amplitude based on spinor techniques which produce expressions for amplitudes that are perfectly suited for their implementation in numerical simulations. Elements of impact studies, including predictions for experiments at JLab12, JLab20+ and the Electron-Ion Collider (EIC), are studied by means of the PARTONS software and the EpIC Monte Carlo event generator.
\end{Abstract}
\vfill
\begin{Presented}
DIS2023: XXX International Workshop on Deep-Inelastic Scattering and
Related Subjects, \\
Michigan State University, USA, 27-31 March 2023 \\
     \includegraphics[width=9cm]{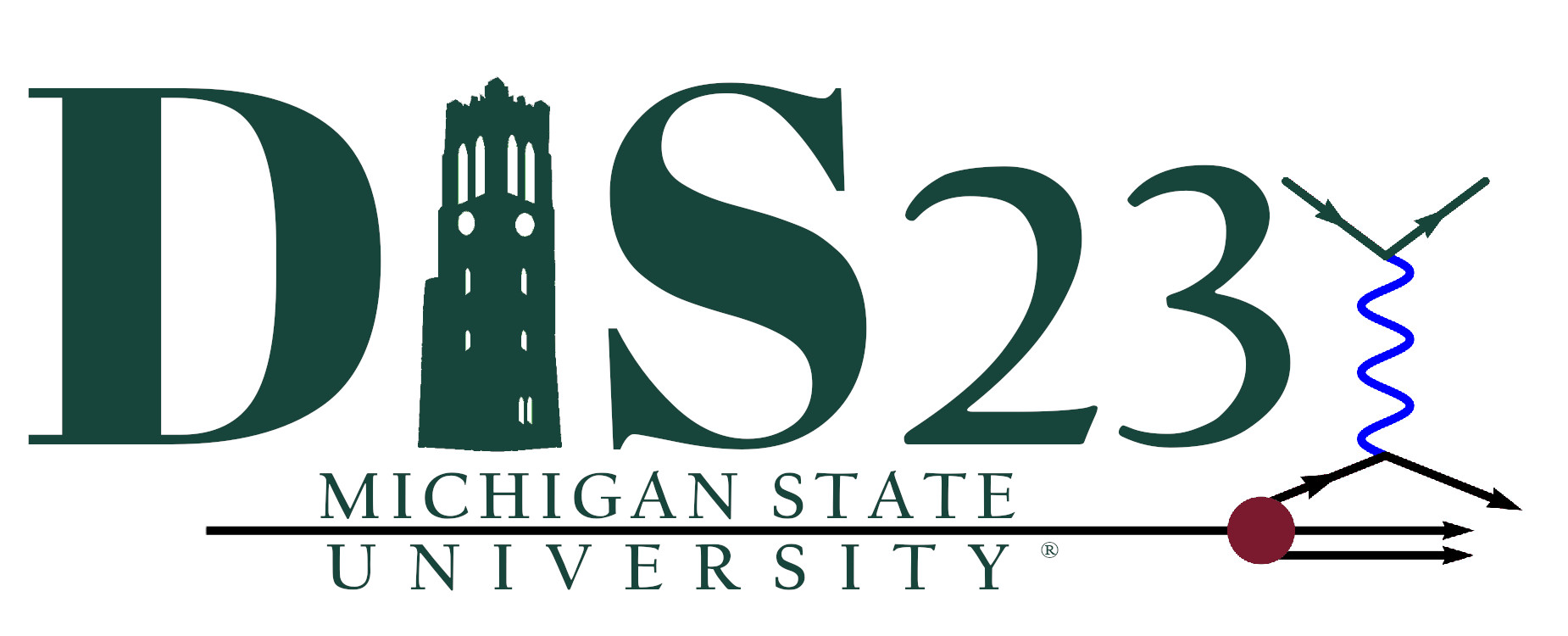}
\end{Presented}
\vfill
\end{titlepage}

%%%%%%%%%%%%%%%%%%%%%%%% 
%   BEGINING OF TEXT   %
%%%%%%%%%%%%%%%%%%%%%%%%

\newgeometry{left=2.5cm,top=2.5cm,right=2cm}

\section{Introduction}
\begin{figure}[htb]
\centerline{
\includegraphics[scale=0.8]{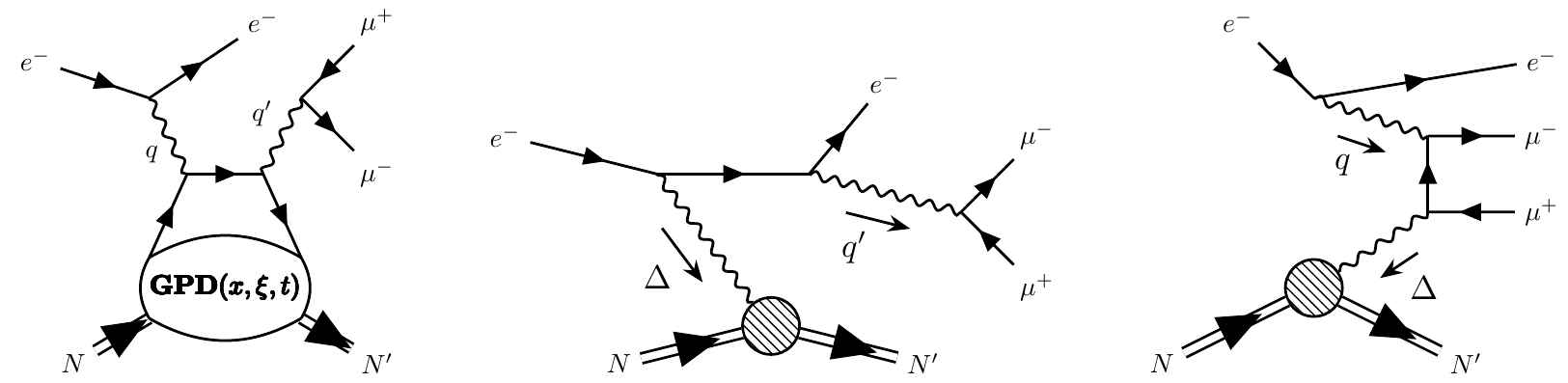}}
\caption{\scriptsize DDVCS (left) and BH diagrams  denoted as BH1 (middle) and BH2 (right). Crossed-counterparts are not included.}
\label{fig:ddvcs_and_bh}
\end{figure}

Generalized parton distributions (GPDs) \cite{Diehl:2003ny,Belitsky:2005qn} are off-forward matrix elements of quark and gluon operators that represent a 3D version of the usual parton distribution functions (PDFs). While PDFs are accessible in inclusive processes GPDs appear in exclusive processes such as deeply virtual and timelike Compton scattering and double deeply virtual Compton scattering (DDVCS).
To access GPDs in DDVCS \cite{Belitsky:2002tf, guidal2003} one  considers the exclusive electroproduction of a lepton pair,
\begin{equation}
    e(k) + N(p) \to e'(k') + N'(p') + \mu^+(\ell_+) + \mu^-(\ell_-) \,,
    \label{reaction}
\end{equation}
in the generalized Bjorken regime where at least one of the two virtualities $Q^2 = -q^2 = -(k-k')^2$ and $Q^{\prime 2} = q^{\prime 2} = (\ell_+ + \ell_-)^2$ is large. This process  receives contributions not only from pure DDVCS, but also from a QED background known as the Bethe-Heitler (BH) sub-process, vid.~Fig.~\ref{fig:ddvcs_and_bh}.

Present interest in DDVCS is rooted on the possibility of directly accessing GPDs in the region $x\neq\pm\xi$ in a leading order (LO) analysis. This is a consequence of the existence of two virtualities $Q^2 $ and $Q^{\prime 2}$ which modifies, with respect to DVCS and TCS, the coefficient function to be convoluted with the GPDs.  In terms of the skewness $\xi$ and the {\it generalized} Bj{\"o}rken variable $\rho$,
\begin{equation}
    \xi = \frac{-\Delta\bq}{2\bp\bq},\quad \rho = \frac{-\bq^2}{2\bp\bq}\,,
\end{equation}
where $\bp = (p+p')/2$, $\bq = (q+q')/2$ and $\Delta = p'-p$, the DDVCS amplitude depends on the GPDs via the Compton form factors (CFFs):
\begin{equation}\label{A_ddvcs}
    {\rm CFF} \sim {\rm PV}\left( \int_{-1}^1 dx\frac{1}{x - \rho} {\rm GPD}(x, \xi, t)\right) - \int_{-1}^1 dx\ i\pi\delta(x - \rho){\rm GPD}(x, \xi, t) \pm \cdots\,.
\end{equation}
Here, the $+$ ($-$) sign corresponds to axial (vector) GPDs, the ellipses accounts for $x\rightarrow -x$ terms, ${\rm PV}$ stands for Cauchy's principal value and $t = \Delta^2$ is the usual Mandelstam variable. As a result, one can measure GPDs for $x = \rho$ for which $\rho \neq \pm\xi$ as long as both vitualities $Q^2, Q^{\prime 2}$ are non-zero. This is different from the DVCS case\footnote{Or TCS with $\xi\rightarrow -\xi$} for which the CFFs in the amplitude enter as in Eq.~(\ref{A_ddvcs}) with $\rho\rightarrow\xi$, which restricts the LO study of GPDs to the line $x = \xi$.

Although a quite detailed study of the phenomenological peculiarities of DDVCS already exists \cite{Belitsky:2003fj}, we revisit this process as it will be considered in upcoming experiments \cite{Chen:2014psa,Camsonne:2017yux,Zhao:2021zsm, AbdulKhalek:2021gbh, Anderle:2021wcy}. For this purpose we \cite{Deja:2023ahc} make use of Kleiss-Stirling (KS) techniques \cite{Kleiss:1984dp, Kleiss:1985yh}, which deals directly with the amplitude and render expressions that are perfectly suited for implementation in PARTONS platform \cite{Berthou:2015oaw} and so for phenomenological studies.

\section{Formulation {\`a} la Kleiss-Stirling}

In 1980s, Kleiss and Stirling developed some spinor techniques to compute  scattering amplitudes as an alternative to the usual approach based on dealing with traces of Dirac-gamma matrices. In that regard, the following products of spinors for two light-like vectors $a$ and $b$ become the building blocks of the amplitudes and define two scalars ($\pm$ stand for helicities):
\begin{align}
    s(a, b) & = \bar{u}(a,+)u(b,-) = -s(b, a)\,, \label{sKS_def}\\
    t(a, b) & = \bar{u}(a,-)u(b,+) = [s(b, a)]^*\,. \label{tKS_def}
\end{align}

Explicit computation of these bilinears show that $s(a, b)$ acquires the simple form:
\begin{equation}\label{sKS_expression}
    s(a, b) = (a^2 + ia^3)\sqrt{\frac{b^0 - b^1}{a^0 - a^1}} - (a\leftrightarrow b)\,,
\end{equation}
as long as $a\cdot\kappa_0\neq 0$ and $b\cdot\kappa_0 \neq 0$ with $\kappa^\mu_0 = (1, 1, 0, 0)$.

In turn, we can define two functions that will play a key role on the computation: the contraction of two currents
\begin{align}\label{function_f}
    f(\lambda, k_0, k_1; \lambda', k_2, k_3) = & \bar{u}(k_0,\lambda)\gamma^\mu u(k_1, \lambda)\bar{u}(k_2,\lambda')\gamma_\mu u(k_3,\lambda') \nonumber\\
    = & 2 [ s(k_2,k_1)t(k_0,k_3)\delta_{\lambda-}\delta_{\lambda'+} + t(k_2,k_1)s(k_0,k_3)\delta_{\lambda+}\delta_{\lambda'-} \nonumber\\
    & + s(k_2,k_0)t(k_1,k_3)\delta_{\lambda+}\delta_{\lambda'+} + t(k_2,k_0)s(k_1,k_3)\delta_{\lambda-}\delta_{\lambda'-} ]\,,
\end{align}
and the contraction of a current with a light-like vector $a$:
\begin{equation}\label{function_g}
    g(s, \ell, a, k) = \bar{u}(\ell, s)\slashed{a}u(k, s) 
    = \delta_{s+}s(\ell,a)t(a,k) + \delta_{s-}t(\ell,a)s(a,k)\,.
\end{equation}

\subsection{Example: DDVCS sub-process {\`a} la KS}
Making use of the quantities defined above, for the case of the left diagram in Fig.~\ref{fig:ddvcs_and_bh} which corresponds to the pure DDVCS contribution, the amplitude may be written as:
\begin{equation}\label{iMddvcs}
    i\M_{\rm DDVCS} = \frac{-ie^4}{(Q^2-i0)(Q'^2+i0)}\left( i\M^{(V)}_{\rm DDVCS} + i\M^{(A)}_{\rm DDVCS} \right)\,,
\end{equation}
where $i\M^{(V)}_{\rm DDVCS}$ and $i\M^{(A)}_{\rm DDVCS}$ correspond to the vector and the axial contributions to the amplitude, respectively. They read:
\begin{align}\label{iMVddvcs}
    i\M^{(V)}_{\rm DDVCS} = & -\Bigg[ f(s_\ell, \ell_-, \ell_+; s, k', k) -  g(s_\ell,\ell_-,n^\star,\ell_+)g(s, k',n,k) - g(s_\ell,\ell_-,n,\ell_+)g(s, k',n^\star,k) \Bigg] \nonumber\\
    & \times \frac{1}{2} \Bigg[ (\cffh + \cffe) [ Y_{s_2s_1}g(+,r'_{s_2},n,r_{s_1}) + Z_{s_2s_1}g(-,r'_{-s_2},n,r_{-s_1}) ] - \frac{\cffe}{M} \Jt_{s_2s_1} \Bigg]\,,
\end{align}
and
\begin{equation}\label{iMAddvcs}
    i\M^{(A)}_{\rm DDVCS} = \frac{-i}{2} \epsilon^{\mu\nu}_\perp j_\mu(s_\ell,\ell_-,\ell_+)j_\nu(s, k', k)\left[ \cffht\Jofplus_{s_2s_1} + \cffet\frac{\Delta^+}{2M}\Jtfplus_{s_2s_1} \right]\,.
\end{equation}

In these sub-amplitudes, $M$ is the target mass, $f$ and $g$ are given in Eqs.~(\ref{function_f}) and (\ref{function_g}), and $\Jt, \Jofplus, \Jtfplus, Y, Z$ are combinations of scalars in Eqs.~(\ref{sKS_def}) and (\ref{tKS_def}) dependent on the spin and momentum of the target in its final ($s_2, p'$) and initial ($s_1, p$) states. $j_\mu$ stands for the lepton currents. Finally, $n$ and $n^\star$ are vectors defining the ``plus'' and ``minus'' light-cone directions as considered in \cite{Belitsky_2000, Belitsky:2003fj, Deja:2023ahc}. This formulation, including BH contributions, was numerically validated against DVCS and TCS limits \cite{Deja:2023ahc}.

\section{DDVCS observables}
In this section we present selected DDVCS observables in the kinematics of current and future experiments, showing that GPD model dependence can also be addressed. For this purpose we use the GK \cite{Goloskokov_2007, Goloskokov_2007_2}, VGG \cite{guichon1998, guichon1999, Goeke_2001, guidal2005} and MMS \cite{Mezrag:2013mya} GPD models implemented in PARTONS. For angles referring to the produced lepton pair, denoted with subscript $\ell$, we make use of the BDP frame \cite{Berger:2001xd} which is common in TCS studies. 

The selected observables are unpolarized differential cross-sections (right and left arrows stand for positive and negative helicity of the incoming electron beam, respectively):
\begin{equation}
    \sigma_{UU}(\phiLBDP) = \int_0^{2\pi} d\phi\int_{\pi/4}^{3\pi/4}d\thetaLBDP\ \sin\thetaLBDP \scriptstyle{\left( \frac{d^7\sigma^{\rightarrow}}{dx_B dQ^2 dQ'^2 d|t| d\phi d\OmegaLBDP} + \frac{d^7\sigma^{\leftarrow}}{dx_B dQ^2 dQ'^2 d|t| d\phi d\OmegaLBDP} \right)}\,,
    \label{eq:ass1}
\end{equation}
and their cosine components:
\begin{equation}
\sigma_{UU}^{\cos(n \phiLBDP)}(\phiLBDP) = M_{UU}^{\cos(n \phiLBDP)}\cos(n \phiLBDP) \,,
\end{equation}
out of which $\sigma_{UU}^{\cos\phiLBDP}$ is proportional to the interference between BH and DDVCS, providing information on the real part of the CFFs. From the last equation, with $N=2\pi$ for $n = 0$ and $N=\pi$ for $n > 0$, the corresponding cosine moment reads:
\begin{equation}
M_{UU}^{\cos(n \phiLBDP)} = \frac{1}{N}\int_0^{2\pi} d\phiLBDP \cos(n \phiLBDP)  \sigma_{UU}(\phiLBDP)  \,.
\end{equation}

We also consider asymmetries for a longitudinally polarized electron beam:
\begin{equation}
    A_{LU}(\phiLBDP) = \Delta\sigma_{LU}(\phiLBDP) / \sigma_{UU}(\phiLBDP)\,,
    \label{eq:ass2}
\end{equation}
where $\Delta\sigma_{LU}$ corresponds to the same integral as (\ref{eq:ass1}) upon the change $(d^7\sigma^{\rightarrow} + d^7\sigma^{\leftarrow}) \rightarrow (d^7\sigma^{\rightarrow} - d^7\sigma^{\leftarrow})$.

\begin{figure}[htb]
    \centering
    \includegraphics[width=0.24\textwidth]{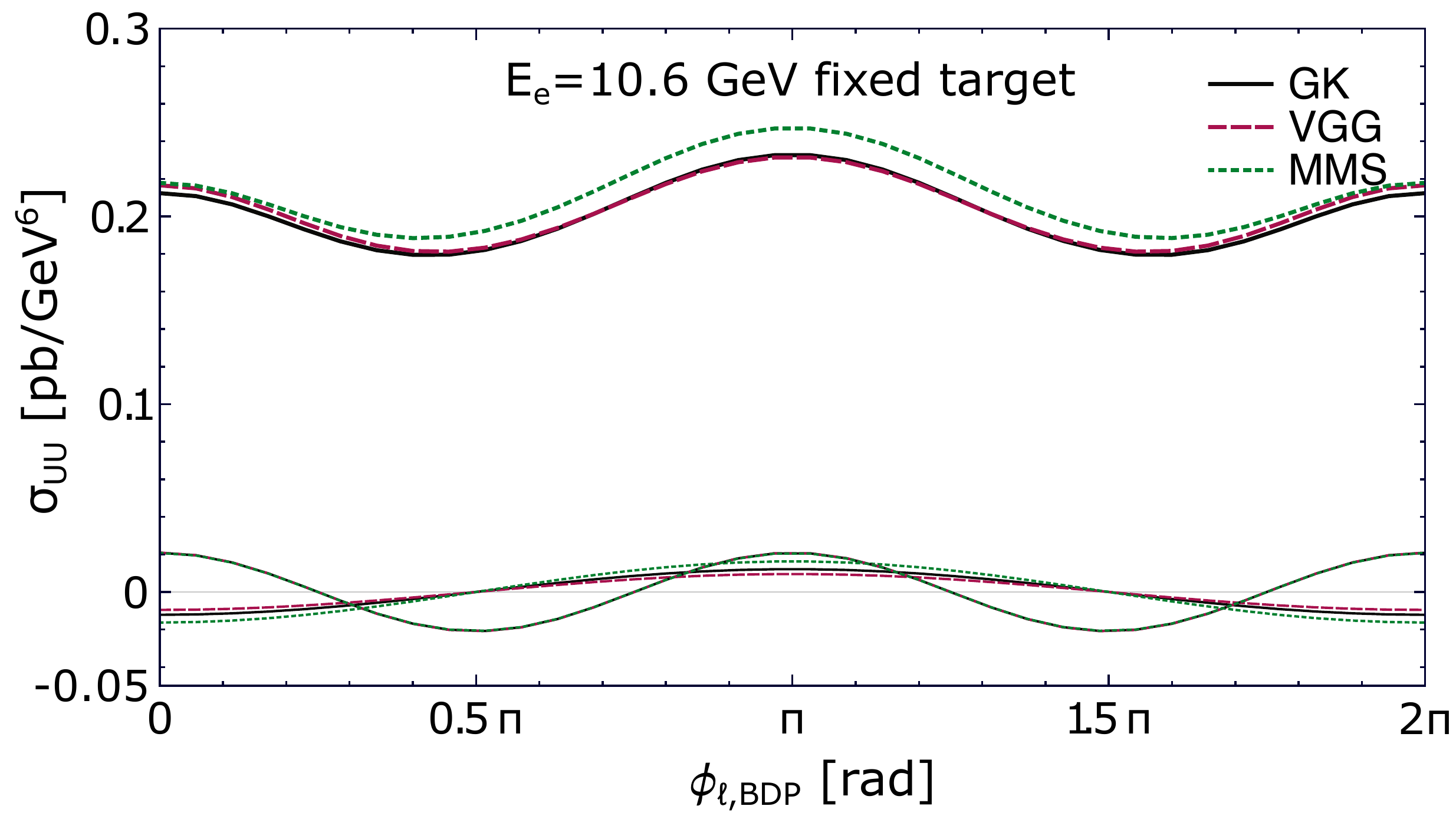}
    \includegraphics[width=0.24\textwidth]{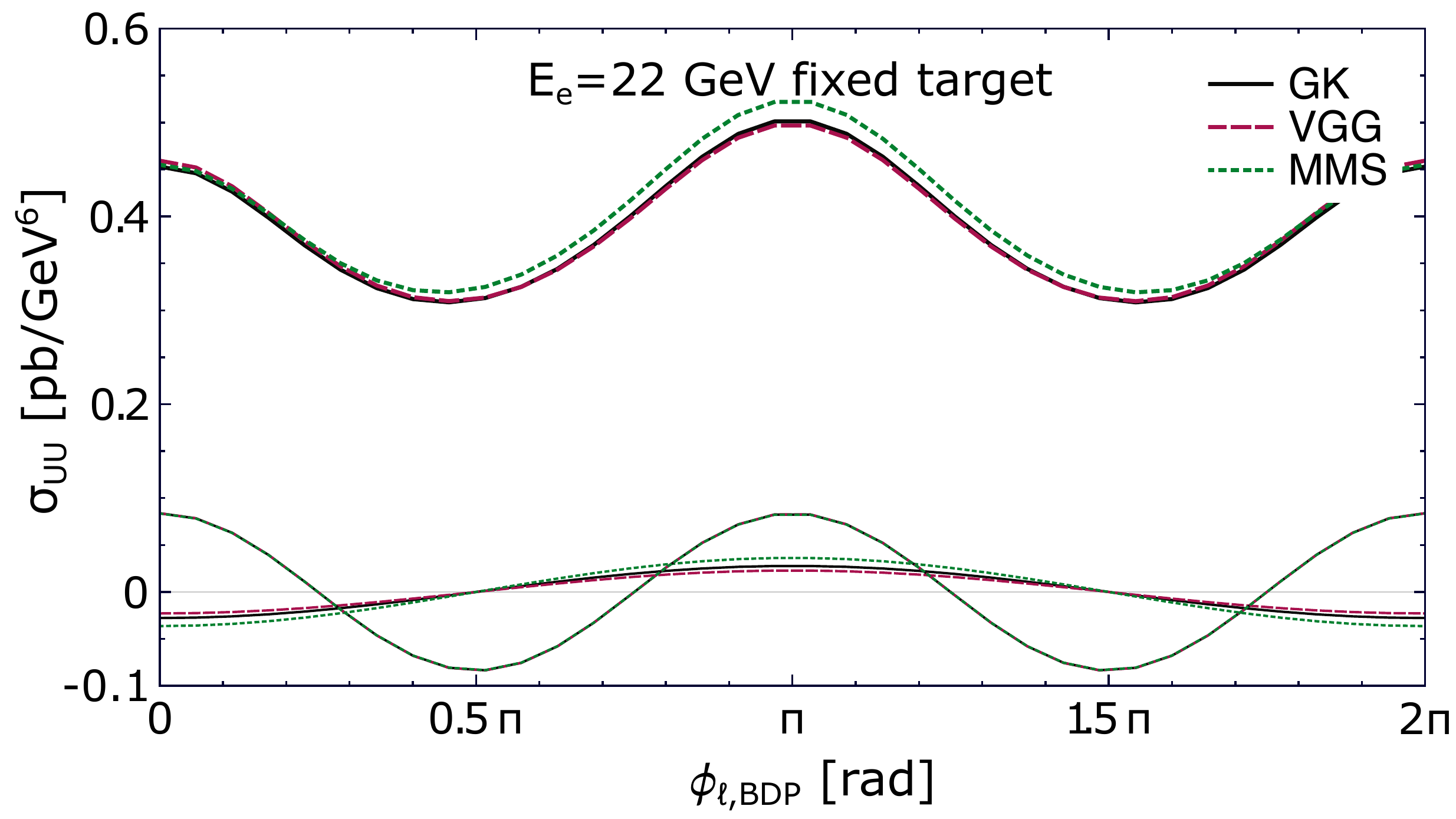}
    \includegraphics[width=0.24\textwidth]{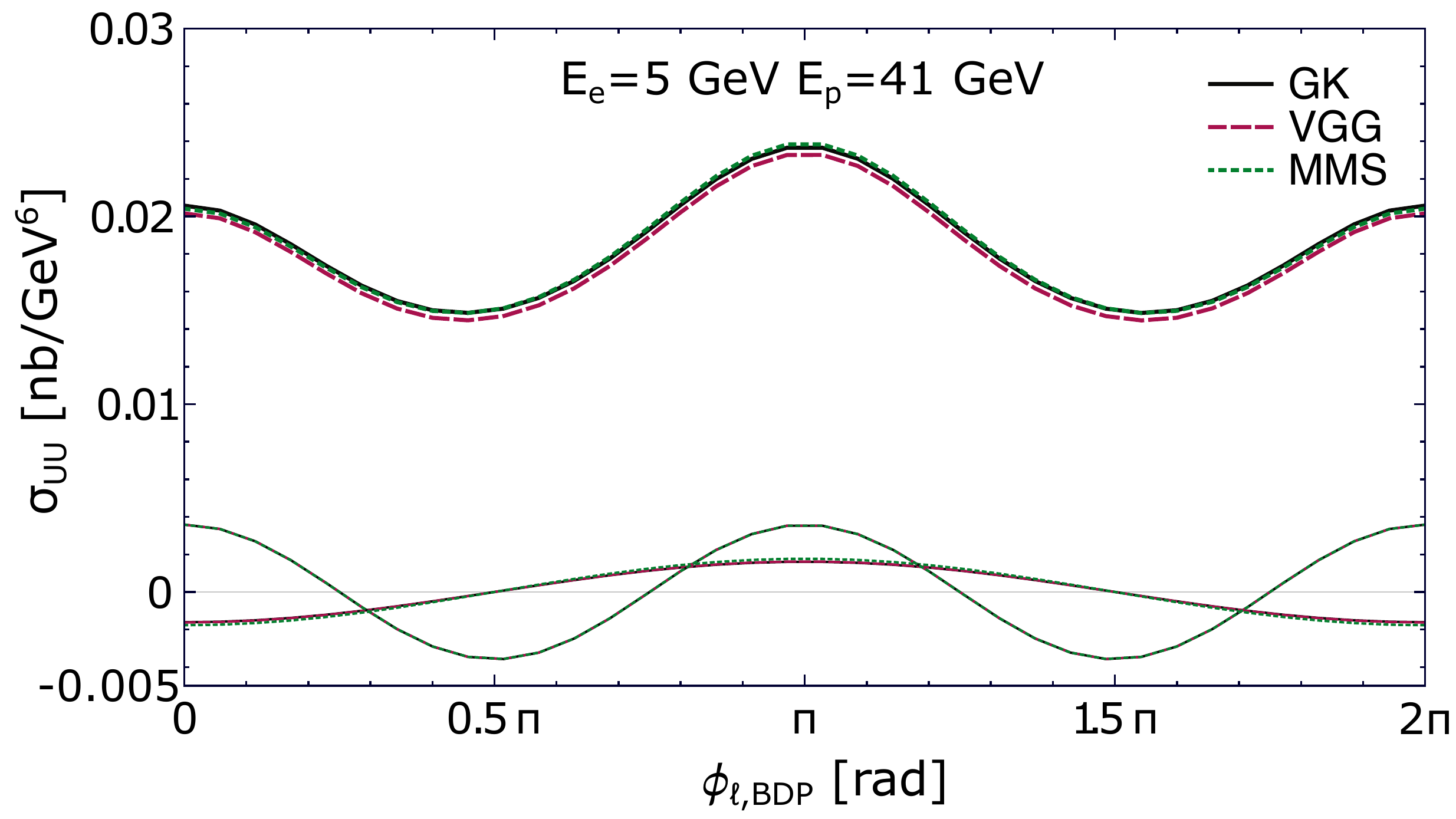}
    \includegraphics[width=0.24\textwidth]{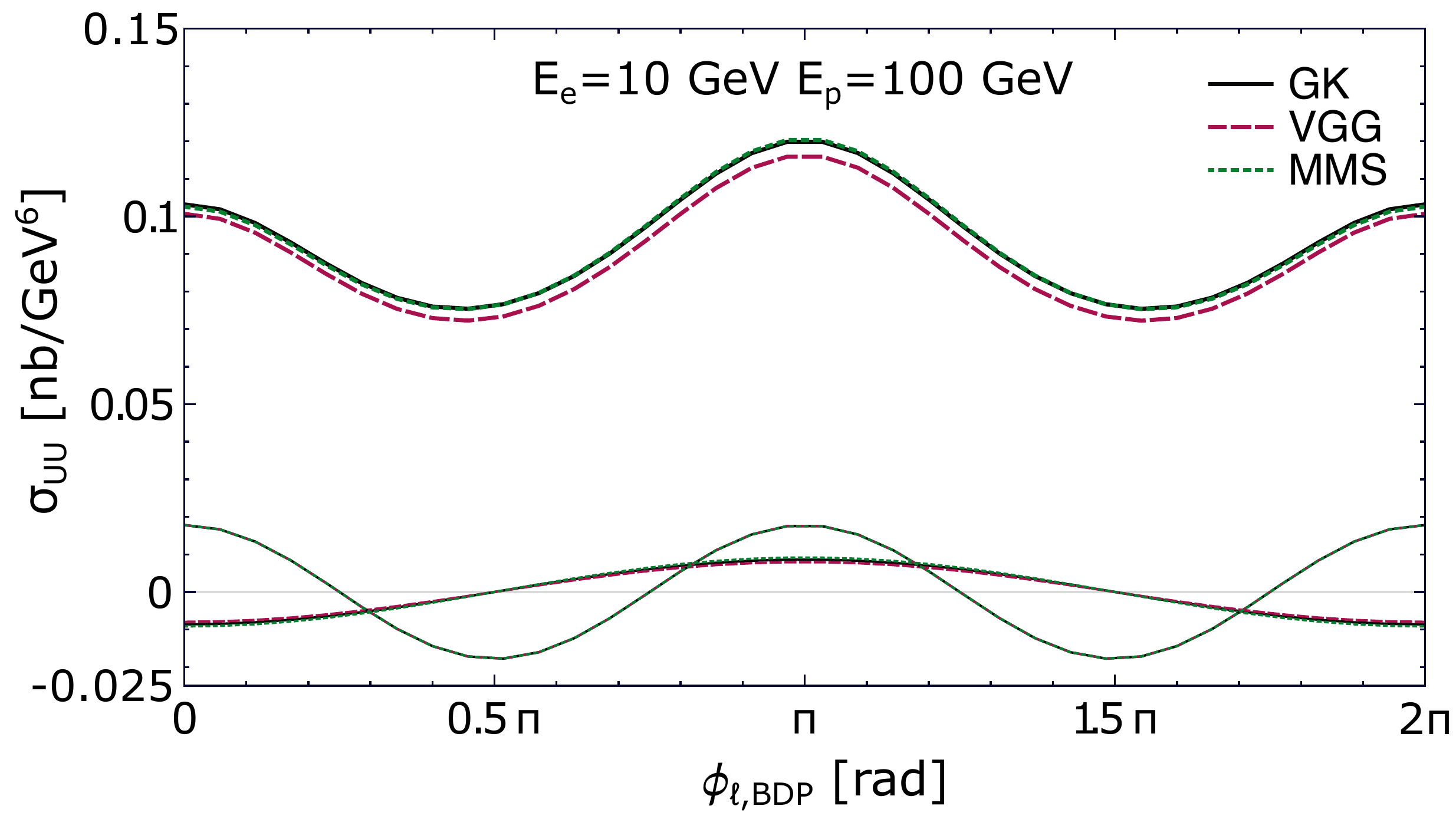}
    \caption{\scriptsize Unpolarized cross-section, $\sigma_{UU}(\phiLBDP)$, and its  $\sigma_{UU}^{\cos\phiLBDP}(\phiLBDP)$ and $\sigma_{UU}^{\cos2\phiLBDP}(\phiLBDP)$ components for beam energies specified in the plots and extra kinematic conditions given in text. From left to right: JLab12, JLab20+, EIC 5$\times$41 and EIC 10$\times$100.}
    \label{fig:denominators}
\end{figure}

\begin{figure}[htb]
    \centering
    \includegraphics[width=0.24\textwidth]{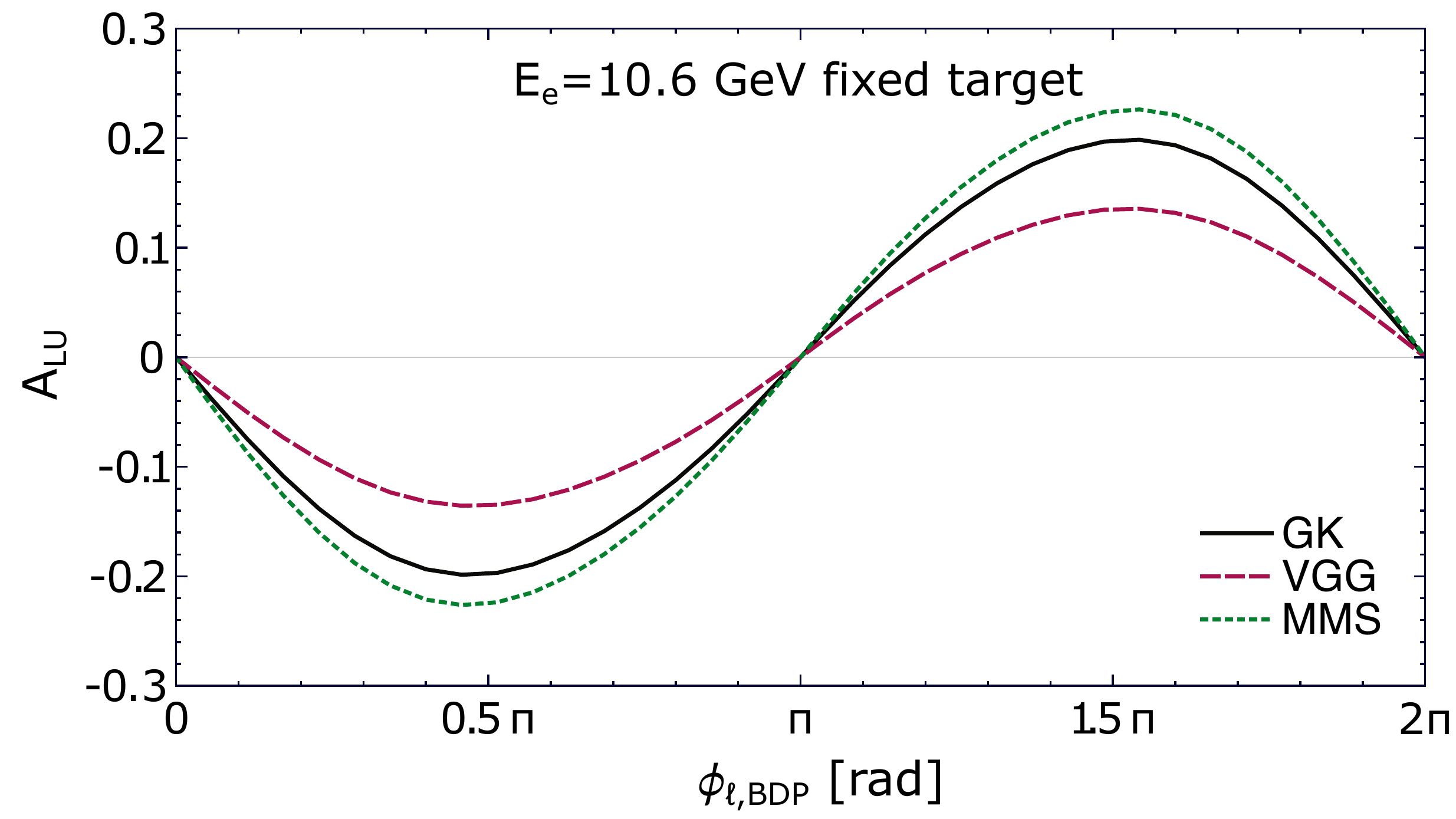}
    \includegraphics[width=0.24\textwidth]{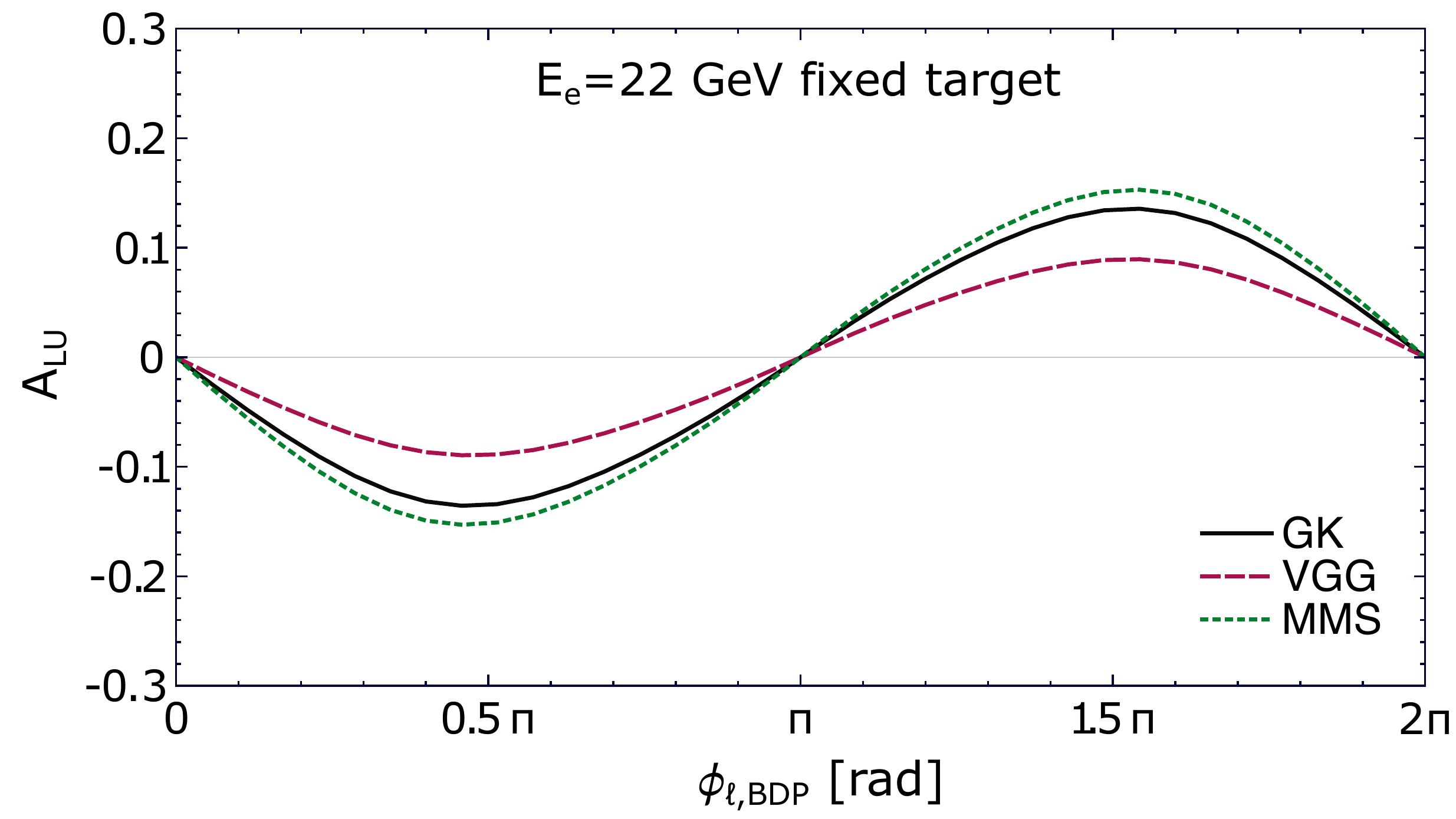}
    \includegraphics[width=0.24\textwidth]{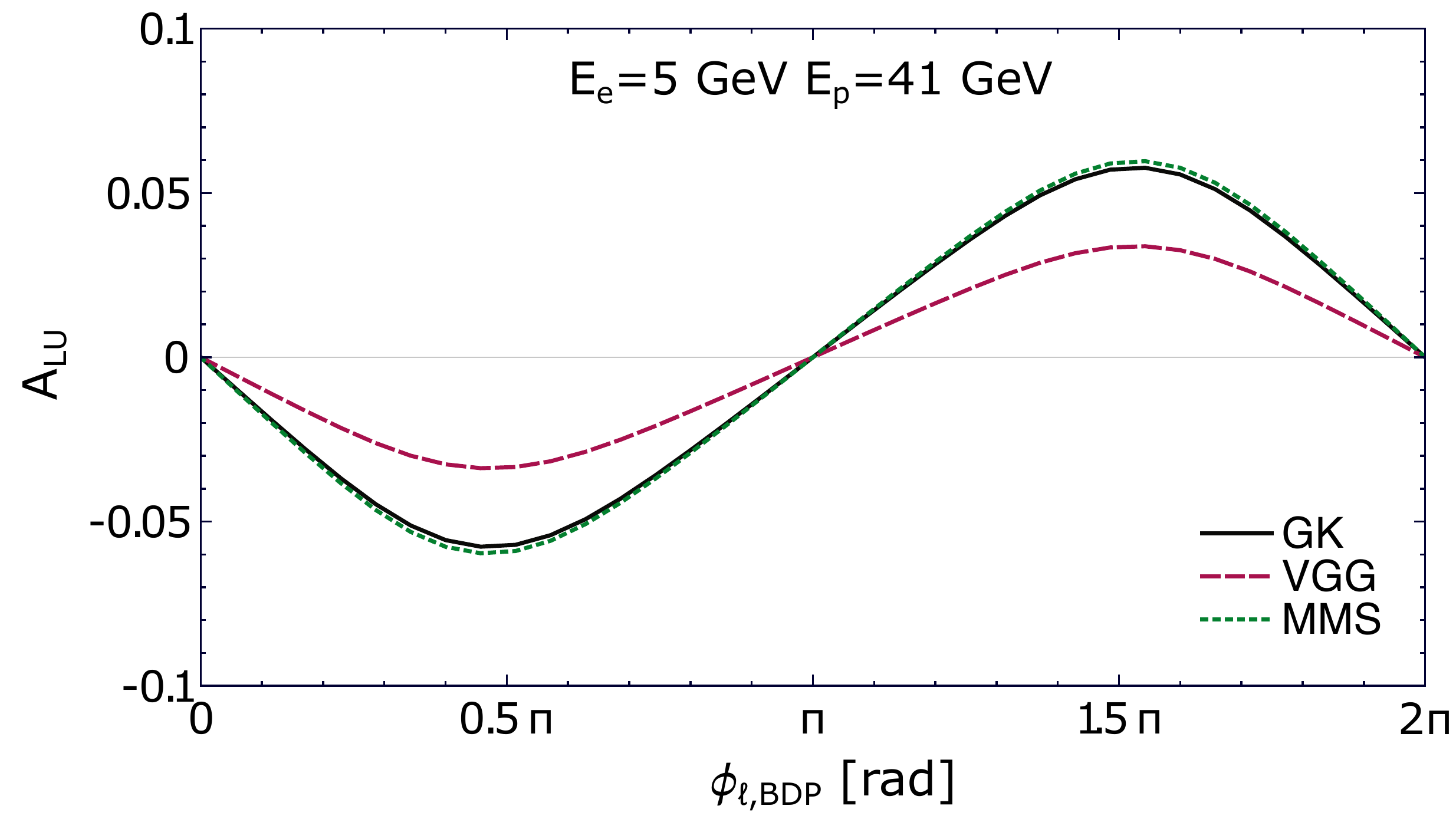}
    \includegraphics[width=0.24\textwidth]{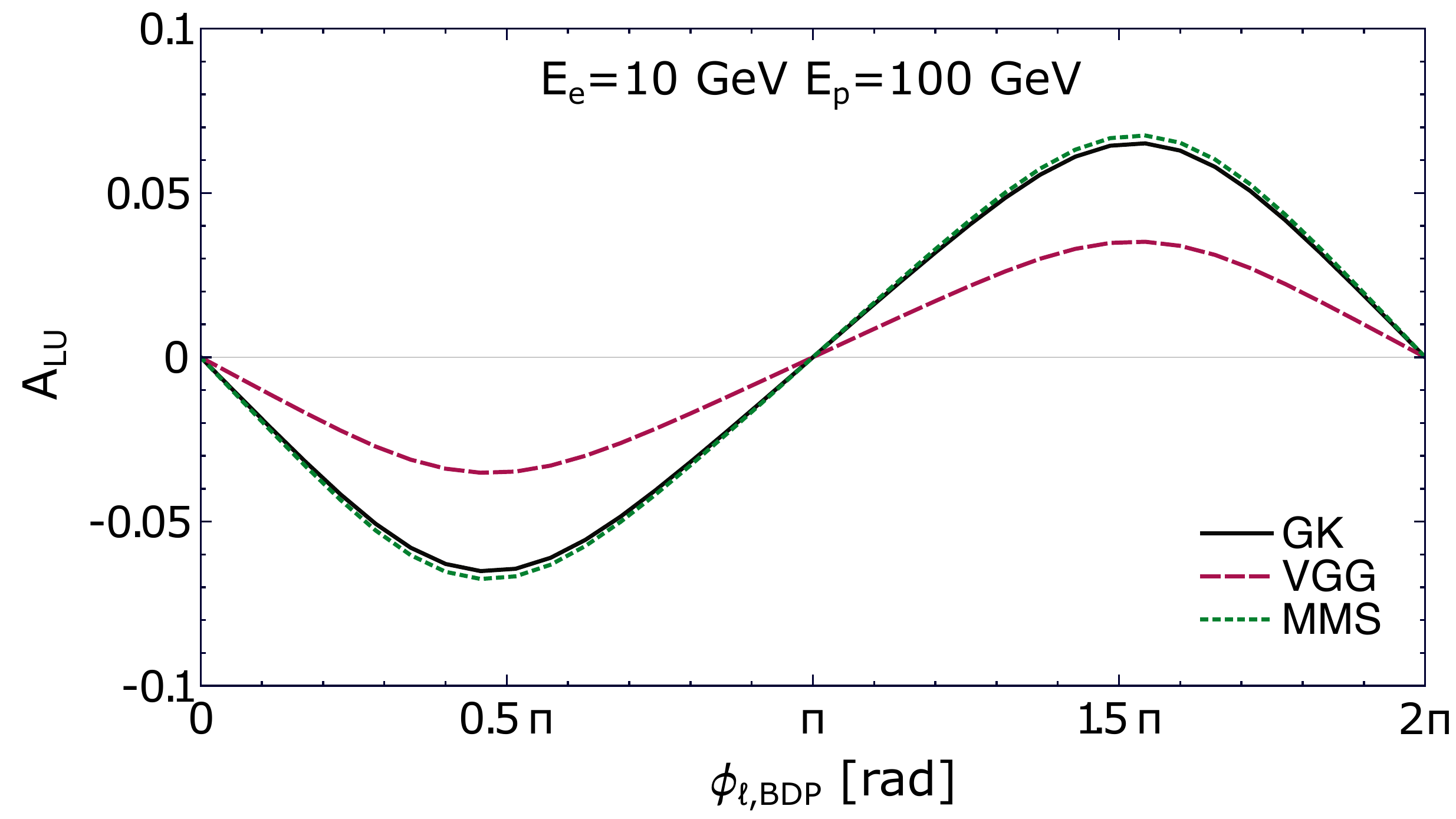}
    \caption{\scriptsize Asymmetry $A_{LU}(\phiLBDP)$ for beam energies specified in the plots and extra kinematic conditions given in text. From left to right: JLab12, JLab20+, EIC 5$\times$41 and EIC 10$\times$100.}
    \label{fig:asymetries}
\end{figure}

The predictions for JLab12, JLab20+ and EIC experiments (for two configurations of beam energies) are shown in Fig.~\ref{fig:denominators} for unpolarized cross-sections and their cosine components, and in Fig.~\ref{fig:asymetries} for the asymmetries. For JLab, $t = -0.2$ GeV$^2$ and in particular $y = 0.5$ for JLab12, while $y = 0.3$ for JLab20+. For both configurations of EIC: $t = -0.1$ GeV$^2$, $y = 0.15$. In all experiments, $Q^2 = 0.6$ GeV$^2$ and $Q^{\prime 2} = 2.5$ GeV$^2$.


\section{Monte Carlo study}
The cross-section formulae we obtained have been implemented in the open-source PARTONS framework. The implementation of DDVCS in the EpIC Monte Carlo (MC) generator \cite{Aschenauer:2022aeb} has followed, making our work directly applicable for experimental analysis. Our results do not include any detector effects. 

\begin{figure}[htb]
    \centering
    \includegraphics[width=0.24\textwidth]{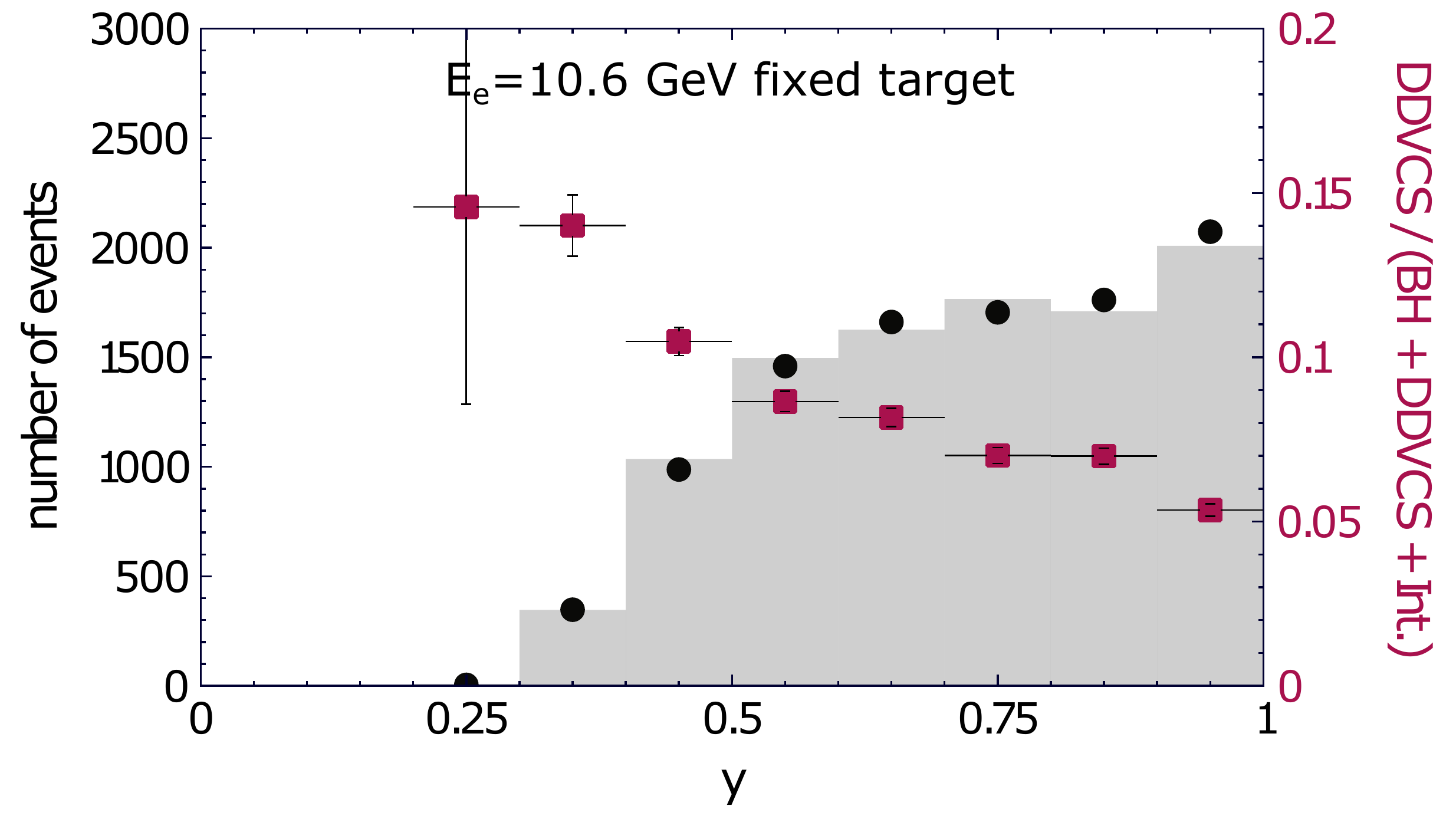}
    \includegraphics[width=0.24\textwidth]{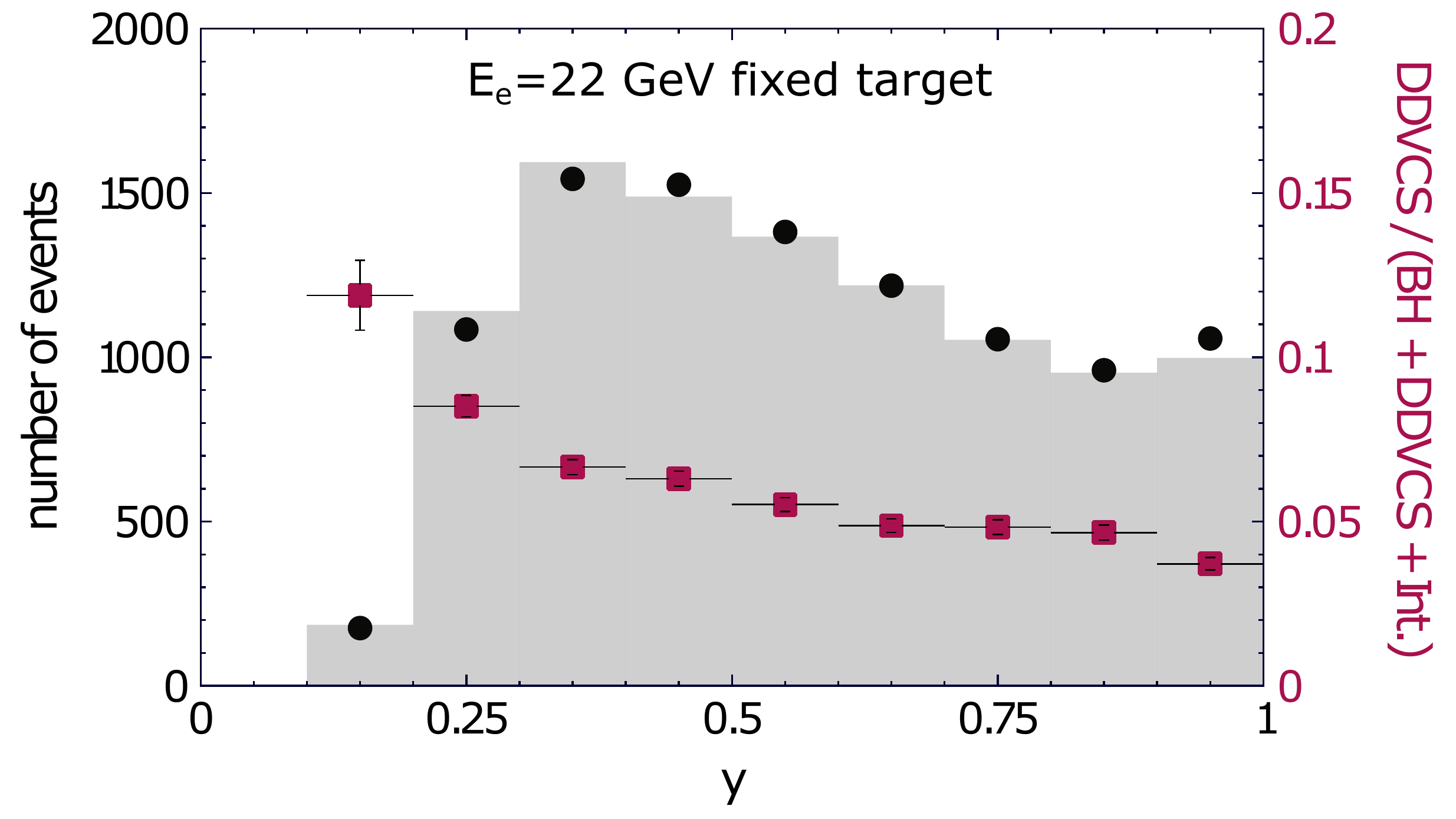}
    \includegraphics[width=0.24\textwidth]{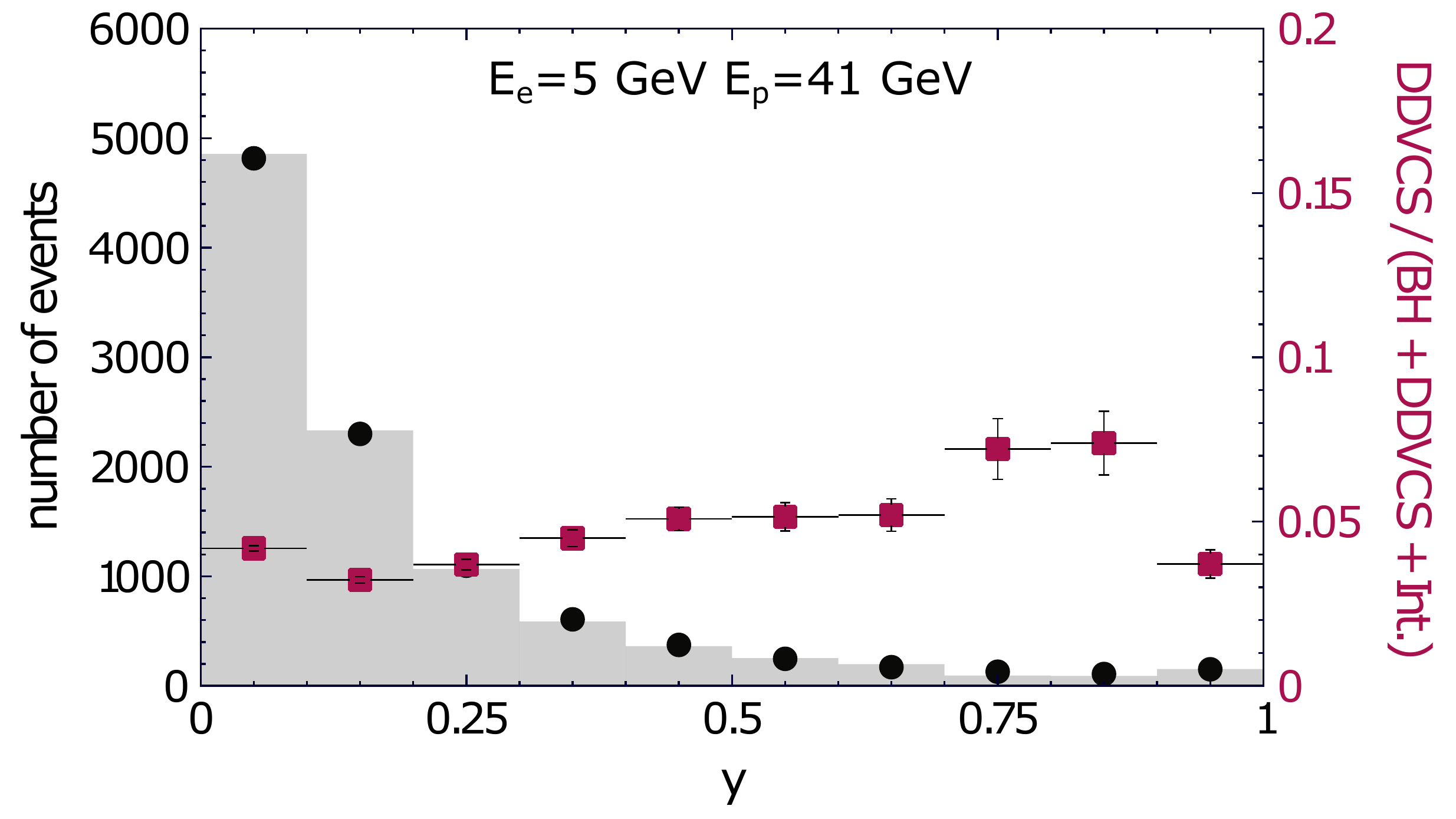}
    \includegraphics[width=0.24\textwidth]{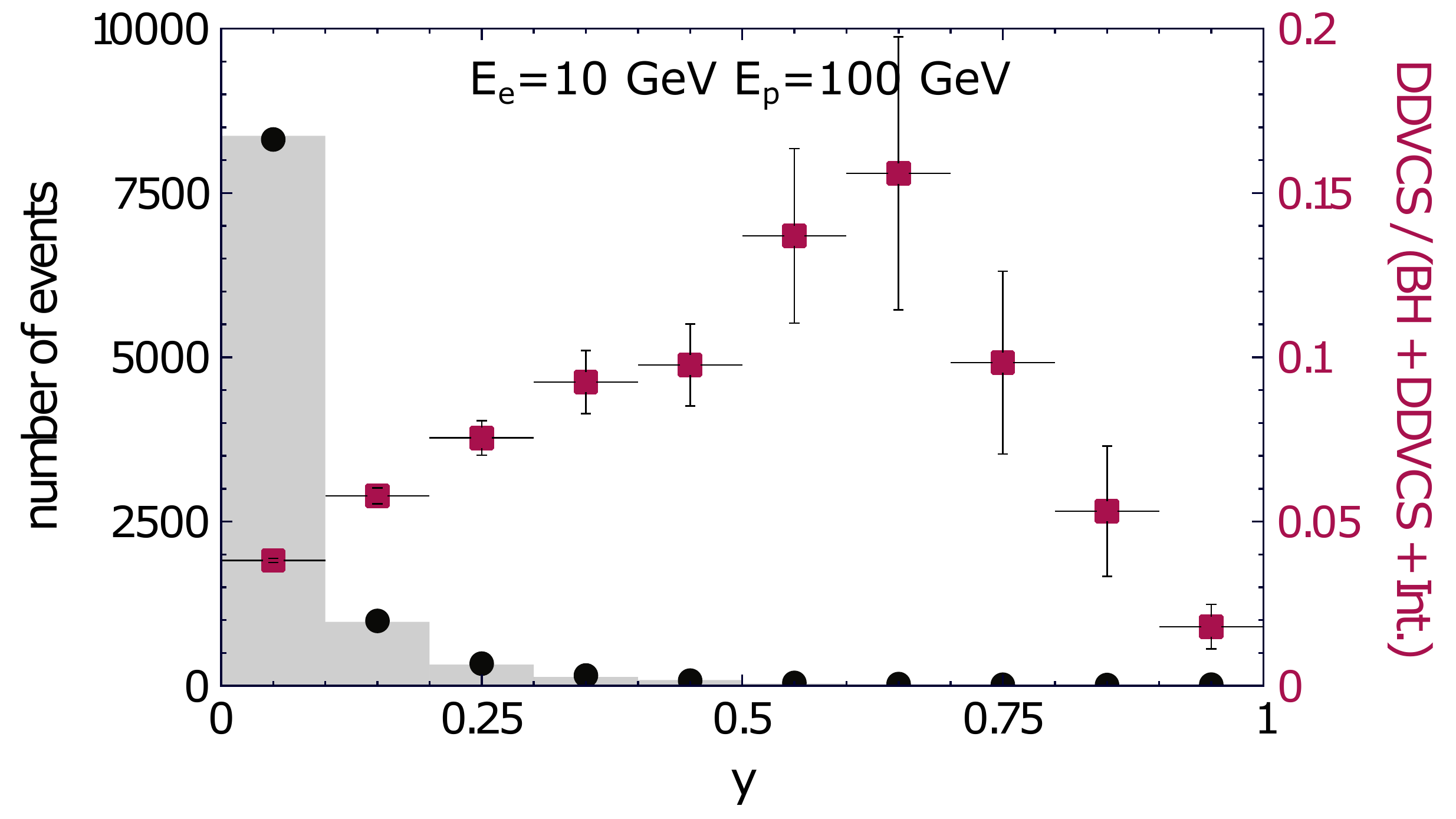}
    \caption{\scriptsize Distributions of Monte Carlo events as a function of the inelasticity variable $y$. Each distribution is populated by 10000 events generated for the beam energies specified in the plots. Extra kinematics indicated in the text. From left to right: JLab12, JLab20+, EIC 5$\times$41 and EIC 10$\times$100.}
    \label{figure:MCHist}
\end{figure}

The distribution of MC events as a function of $y = pq/(pk)$ is shown in Fig.~\ref{figure:MCHist}. The total cross-section for the scattering (\ref{reaction}) is given in Table \ref{tab:MCCS}. Kinematical cuts for the integration are: $y \in (0, 1)$, $Q^2 \in (0.15, 5)\ \mathrm{GeV}^2$, $Q^{\prime 2} \in (2.25, 9)\ \mathrm{GeV}^2$, $\phi, \phiL \in(0.1, 2\pi - 0.1)$ rad, $\thetaL \in (\pi/4, 3\pi/4)$ rad and $|t|\in(0.1, 0.8)$ GeV$^2$ for JLab and $(0.05, 1)$ GeV$^2$ for EIC. In this table we also specify the integrated luminosity needed to record 10000 events presented in Fig.~\ref{figure:MCHist}, and the fraction of events recovered after considering the lower cut in $y$, namely $y_{\rm min}$ in the table, coming from the experimental constraints in the reconstruction of the scattered electron. In Fig.~\ref{figure:MCHist} we also show the expected number of events, coming from a direct seven-fold integration of cross-section (black dots). The comparison between the obtained values and MC samples (grey bands) proves the correctness of the generation process. In Fig.~\ref{figure:MCHist} we also display the fraction of pure DDVCS sub-process in the sample (red boxes). The smallness of this quantity indicates the need of observables proportional to the interference between DDVCS and BH. 

\begin{table}[htb]
\centering
\scalebox{0.7}{\begin{tabular}{lcccccc}
\hline\hline
& & & & & \\[-10pt]
Experiment & Beam energies & Range of $|t|$ & $\sigma \rvert_{0<y<1}$ & $\mathcal{L}^{10\mathrm{k}}\rvert_{0<y<1}$ & $y_{\mathrm{min}}$ & $\sigma \rvert_{y_{\mathrm{min}} < y < 1} / \sigma \rvert_{0<y<1}$ \\ 
& $[\mathrm{GeV}]$ & $[\mathrm{GeV}^2]$ & $[\mathrm{pb}]$ & $[\mathrm{fb}^{-1}]$ & & \\[5pt] 
JLab12 & $E_{e} = 10.6$, $E_p = M$ & $(0.1, 0.8)$ & $0.14$ & $70$ & $0.1$ & $1$\\
JLab20+ & $E_{e} = 22$, $E_p = M$ & $(0.1, 0.8)$ & $0.46$ & $22$ & $0.1$ & $1$\\ 
EIC & $E_{e} = 5$, $E_p = 41$ & $(0.05, 1)$ & $3.9$ & $2.6$ & $0.05$ & $0.73$\\
EIC & $E_{e} = 10$, $E_p = 100$ & $(0.05, 1)$ & $4.7$ & $2.1$ & $0.05$ & $0.32$ \\
[3pt]
\hline\hline
\end{tabular}}
\caption{\scriptsize Total cross-section for (\ref{reaction}), $\sigma \rvert_{0<y<1}$, obtained for the beam energies indicated in each row, $y\in(0, 1)$ and extra kinematics indicated in the text. Corresponding integrated luminosity required to obtain 10000 events is denoted by $\mathcal{L}^{10\mathrm{k}}\rvert_{0<y<1}$. Fraction of events left after restricting the range of $y$ to $(y_{\mathrm{min}}, 1)$ is given in the last column.}
\label{tab:MCCS}
\end{table}

As a final remark, we conclude that asymmetries at LO are of order 15-20\% for JLab and 3-7\% for EIC, large enough for DDVCS to be considered a relevant part of GPD programmes in current and future experimental facilities. The inclusion of NLO corrections \cite{Pire:2011st} should not change this conclusion.

\scriptsize{{\bf Acknowledgements.} The works of J.W. are  supported by the grant 2017/26/M/ST2/01074 of the National Science Center (NCN), Poland. Development of EpIC Monte Carlo generator by P.S. was supported by the grant 2019/35/D/ST2/00272 of the NCN. This work is also partly supported by the COPIN-IN2P3 and by the European Union’s Horizon 2020 research and innovation programme under grant agreement No 824093. The works of V.M.F.~are supported by PRELUDIUM grant 2021/41/N/ST2/00310 of the NCN.}

%% BIBLIOGRAPHY %%
\bibliographystyle{amsrefs}
\bibliography{eprint.bib}

\end{document}